\documentclass[11pt]{article}
\usepackage[utf8]{inputenc}
\usepackage{amsmath, amssymb}
\usepackage{graphicx}
\usepackage{caption}
\usepackage{hyperref}
\usepackage{authblk}
\usepackage{cite}

\usepackage{subcaption}

\usepackage[ruled, vlined, linesnumbered]{algorithm2e}
\DontPrintSemicolon
\SetKw{KwBy}{by}
\SetKw{KwAnd}{and}
\usepackage{tabto}

\usepackage{placeins}

\newcommand{\subsubruninhead}[1]{{\normalsize\textbf{#1.}}}

\date{}

\title{The Multiple Time-Stepping Method for 3-Body Interactions in High Performance Molecular Dynamics Simulations}
\author[1]{David Martin}
\author[1]{Samuel James Newcome}
\author[1]{Markus Mühlhäußer}
\author[1]{Manish Kumar Mishra}
\author[1]{Fabio Alexander Gratl}
\author[1]{Hans-Joachim Bungartz}
\affil[1]{Chair of Scientific Computing in Computer Science, Department of Computer Science, Technical University of Munich, Boltzmannstr. 3, 85748, Garching bei Muenchen, Germany;\newline
david.martin@in.tum.de}

\begin{document}

\maketitle

\begin{abstract}
Understanding the complex behavior of molecular systems is fundamental to fields such as physics, materials science, and biology. Molecular dynamics (MD) simulations are crucial tools for studying atomic-level dynamics. This work focuses on improving the efficiency of MD simulations involving two-body and three-body interactions. Traditional two-body potentials often can not fully capture the complexity of molecular systems, making the inclusion of three-body interactions important. However, these interactions are in a cubic complexity class, compared to a quadratic one for two-body interactions, and therefore are computationally expensive, even when a cutoff distance is applied. One way to improve efficiency is to use the r-RESPA multiple time-stepping algorithm to reduce the number of three-body interaction calculations. In this work, we investigate this method in the context of High Performance Computing (HPC) methods that parallelize the calculations. In particular, we investigate a communication-reducing distributed-memory parallel method from literature and present a novel shared-memory parallel cutoff method, implemented in the particle simulation library AutoPas. The results and methods are discussed, providing insights into potential advancements in MD simulation efficiency.
\end{abstract}

\section{Introduction}
\label{sec:intro}
Molecular dynamics (MD) simulations are essential tools in physics, materials science, and biology, providing insights into atomic and molecular behavior over time. These simulations are crucial for understanding thermodynamic properties, phase transitions in materials, and biological processes such as protein folding and molecular interactions \cite{Frenkel2002UnderstandingMD,Allen2017ComputerSimulation,Karplus2005ProteinDynamics}. Here, we refer to molecules or atoms as \textit{particles}.

Accurate MD simulations rely on comprehensive modeling of inter-atomic interactions. Two-body potentials, such as Lennard-Jones and Coulomb models, describe the pairwise interactions between particles, where the total force is the sum of forces between all pairs of molecules. However, for complex systems, three-body interactions are often necessary to capture critical behaviors, such as stabilizing specific molecular configurations or more accurate phase transitions \cite{marcelli_molecular_1999,wang_three-body_2006}. While these interactions provide additional accuracy, they are computationally expensive, scaling with $\mathcal{O}(n^3)$ compared to $\mathcal{O}(n^2)$ for two-body interactions. Examples of systems benefiting from three-body potentials include water models, where the angular dependencies play a crucial role, and solid-state systems, where three-body forces help capture lattice stability \cite{Stillinger1974ThreeBodyWater,Cleri1993SolidStability}.

Another computational challenge in MD simulations arises from the need for very small time-steps typically in the femtosecond-range for accurate time integration, ensuring numerical stability and precision in modeling high-frequency molecular vibrations. 

It is unrealistic to calculate large-scale simulations in an acceptable time on standard computers. For these reasons, it is necessary to develop efficient algorithms for High Performance Computing (HPC) systems that parallelize the calculation of particle interactions and make optimal use of the capacities of such compute clusters.

However, three-body interactions remain significantly more expensive to calculate than two-body interactions, underscoring the need for additional optimization techniques to improve computational efficiency. 

One promising approach is the reversible Reference System Propagator Algorithm (r-RESPA), a Multiple Time-Stepping (MTS) technique. This method integrates interactions at different time steps, with high-frequency forces computed more frequently than low-frequency ones, thereby reducing computational cost \cite{griebel_numerical_2007}.

This work explores the application of the r-RESPA algorithm in conjunction with state-of-the-art HPC techniques for MD simulations involving two-body and three-body interactions. Our contribution lies in the combined use of r-RESPA and HPC optimizations specifically tailored for three-body interactions. We have developed a novel three-body cutoff algorithm for shared memory implemented in AutoPas \cite{gratl_n_2022}, which we evaluate in conjunction with the r-RESPA method for simulations involving both two-body and three-body interactions. Additionally, we assess a state-of-the-art distributed memory algorithm from the literature \cite{koanantakool_computation-_2014} with r-RESPA for two-body and three-body simulations, further enhancing the performance and scalability of these simulations on large-scale HPC systems.

The underlying idea is that three-body interactions often act as corrective influences on two-body forces, allowing for larger time steps without compromising simulation accuracy. For instance, studies have shown that three-body interactions are important as they refine the potential energy and equilibrium properties, capturing nuanced physical behaviors. However, they play a secondary role compared to the dominant two-body forces \cite{wang_influence_2006}.

The paper is structured as follows: In section 2, we give a brief overview of Molecular Dynamics Simulations, the numerical methods for computation of MD simulations including MTS methods, as well as practical algorithms from the HPC domain. In section 3 we give a short literature review on how MTS methods have been used so far. In Section 4, we present our implementation and results. Here we investigate different test scenarios with different step-size-factors for three-body interactions.

\section{Background}
\label{sec:background}

To simulate particle motion over a time interval $I = [t_{start}, t_{end}]$, we solve the classical mechanics equations of motion, represented by Newton's equation:

\begin{equation}
F = m \cdot a
\end{equation}

where $m$ is the particle mass and $a$ its acceleration.

Since we can not solve these equations analytically over the whole time interval, we discretize $I$ into time steps of size $\delta t$. At each step, we use numerical methods to solve the equations of motion.

To find the acceleration $a$, we compute the force on each particle at each time step. Using this acceleration, we can calculate velocity and update particle positions. The force is determined by the negative gradient of the overall potential $\Phi$:

\begin{equation}
F = -\nabla \Phi
\end{equation}

To approximate the forces on particles accurately, MD simulations use potential energy functions $\phi$ for different interaction types. These functions can for example be classified by the number of particles involved, with pairwise (two-body) interactions being the most common.

The exact force on a particle $i$ is calculated by summing all potentials (single-body to N-body) and taking the negative gradient:

\begin{equation}
F_i = - \nabla \Phi = - \nabla \left[ \phi_1(i) + \sum_{j}{\phi_2(i,j)} + \sum_{j,k}{\phi_3(i,j,k)} + \dots \right]
\end{equation}

Most of the interaction can be modeled with just the pairwise potential, with subsequent terms being used to apply corrections if needed. However, as more particles are included in the potential, the computational cost increases. This often limits calculations to two-body potentials. Nevertheless, as previously mentioned, the inclusion of three-body forces has been shown to yield more accurate results in certain applications, making them necessary despite the added computational expense \cite{marcelli_molecular_1999}.

\subsection{Potentials}
A common two-body short-range potential used in MD simulations to model electrostatically neutral intermolecular interactions is the \textbf{Lennard-Jones 12-6} (LJ) potential, defined as:

\[
\phi_{2}(i,j) = 4 \epsilon_{LJ} \left[ \left( \frac{\sigma_{LJ}}{r_{ij}} \right)^{12} - \left( \frac{\sigma_{LJ}}{r_{ij}} \right)^{6} \right]
\]

where \( r_{ij} \) is the distance between particles \( i \) and \( j \), and \( \epsilon_{LJ} \) and \( \sigma_{LJ} \) are material-specific parameters.
\bigbreak

The \textbf{Axilrod-Teller-Muto} (ATM) potential is a three-body potential often used in conjunction with the Lennard-Jones potential to model three-body interactions. \cite{axilrod_interaction_1943} It is defined as:

\[
\phi_{3}(i,j,k)= \nu_{ATM} \left[ \frac{1+3 \cos(\theta_i) \cos(\theta_j) \cos(\theta_k)}{(r_{ij} \cdot r_{ik} \cdot r_{jk})^3} \right]
\]

where \( r_{ij} \) is the distance between particles \( i \) and \( j \), \( \theta_i \) is the angle between vectors \( \vec{ij} \) and \( \vec{ik} \), and \( \nu_{ATM} \) is a positive, material-dependent coefficient.

\subsection{Methods for MD Simulations on HPC Systems}
MD simulations involve calculating interactions between up-to a trillion particles \cite{tchipev_twetris_2019} over extended time periods. HPC systems handle this by distributing the computational load across multiple processors, significantly reducing calculation time and allowing for more detailed, longer simulations than on standard computers. In the following, we briefly describe methods that are used to optimize the calculation time.

\textbf{Cutoff Distance}
In order to save computing time in MD simulations, a cutoff distance $r_c$ is often applied to short-range potentials. As these potentials rapidly converge to zero with increasing particle distance, interactions beyond a certain range contribute only marginally and can be ignored. This approach significantly improves performance and reduces the complexity to $\mathcal{O}(n)$, although it introduces an additional approximation. In contrast, implementations that work without a cutoff distance and calculate all possible interactions between particles are referred to as \textit{DirectSum}.

\textbf{Newton's Third Law of Motion}
This law states that the force $F_{B \leftarrow A}$ exerted by one particle $A$ on another particle $B$ is also exerted in the other direction with the same magnitude: $F_{B \leftarrow A} = -F_{A \leftarrow B}$. In the context of MD simulations and HPC, we can exploit this law to save computation time, as we only need to evaluate the force between two particles once. However, to prevent race conditions in parallel computations, algorithms must be specifically designed to handle this, ensuring that the forces are computed consistently across processors.

\textbf{Domain Decomposition}
In order to efficiently distribute the calculation of particle interactions across several processors, various algorithms have been developed that can be used depending on the purpose of the application.

For example, there are algorithms that subdivide the domain spatially. These are particularly suitable for implementations that use a cutoff distance, as the spatial division already implicitly provides information about the distance between particles. A common algorithm used here is the linked cells algorithm. This algorithm divides the simulation domain into a grid and distributes the particles in their respective cells to processors. To calculate the interactions, various traversal techniques can then be used to process the cells. An exemplary traversal that we use is described in more detail in Section \ref{subsec:implementation}.

Note that for implementations that work without a cutoff distance and calculate all possible interactions between particles, such a domain decomposition is not strictly necessary. Instead, particles can simply be distributed across processors, without any spatial considerations. This approach focuses on distributing the particles themselves, which can still lead to efficient load balancing depending on the system's characteristics.

\subsection{Numerical Integration}
As mentioned at the beginning of Section \ref{sec:background}, the time interval to be simulated is divided into discrete time steps, $\delta t$. In each of these steps, the new positions of the particles are calculated using numerical time integration algorithms. Since this time integration is performed only at discrete steps, it is inherently an approximation. 

Especially in simulations involving a large number of time steps, the resulting particle positions towards the end of the simulation may significantly deviate from the exact results. However, in MD simulations, obtaining the exact positions is often less important than accurately preserving other properties, such as statistical quantities or conserved quantities like the total energy, which can still be determined with high accuracy. \cite{griebel_numerical_2007}

To reduce errors, smaller time steps can be chosen. However, depending on the time integration algorithm, multiple force calculations per step are needed, significantly increasing computational effort. Since force computations account for up to 95\% of the runtime in MD simulations \cite{deuflhard_computational_1999}, selecting efficient integration schemes that minimize these calculations is crucial. One algorithm frequently used in MD simulations is the Störmer Verlet algorithm. It achieves second-order accuracy while requiring only a single force computation per step.

Another key advantage of the Störmer Verlet algorithm is its symplectic nature. In Hamiltonian systems, symplecticity ensures the approximate conservation of system energy over time \cite{griebel_numerical_2007}. This property is essential for accurately capturing intrinsic conservation laws governing many physical systems. 
Consequently, the combination of second-order accuracy, computational efficiency, and symplecticity makes the Störmer Verlet algorithm particularly well-suited for MD simulations.

\subsubsection {Multiple Time-Stepping (MTS) Methods}
\begin{figure}[htb]
  \centering
  \begin{minipage}{0.4\textwidth}
    \captionof{figure}{Visualization of the MTS method for two forces $F_0$ and $F_1$ at different time steps. Here $F_0$ is integrated at intermediate steps $\delta t/4$.}
    \label{fig:mts}
  \end{minipage}
  \hfill
  \begin{minipage}{0.55\textwidth}
    \includegraphics[width=\linewidth]{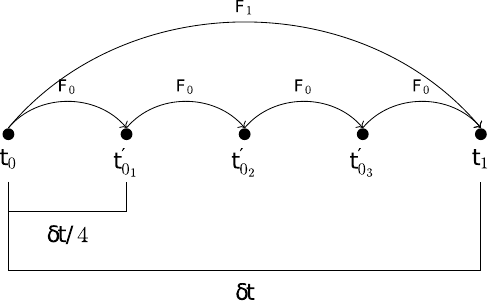}
  \end{minipage}%
\end{figure}

In classical time integration schemes, such as Störmer Verlet, the time step to be selected is based on the highest frequency in the system. These are, for example, bond, angle, and torsion forces. However, other forces, such as non-bonded forces, can have a much lower frequency, and the time steps could be chosen larger without significantly losing accuracy \cite{griebel_numerical_2007}. This is where the Multiple Time-Stepping methods come into play, which decompose the total force into components (e.g., $F_0$ and $F_1$) acting at different frequencies, allowing integration at different time scales. This approach saves calculation time. The time integration for two different forces, such as two-body and three-body forces, can be visualized as shown in Figure \ref{fig:mts}.

\textbf{r-RESPA:} The reversible Reference System Propagation Algorithm (r-RESPA) method or impulse method is a concrete implementation of MTS. In this algorithm, which is based on the classical Störmer Verlet method, different time-scales are used for the different potentials to be calculated. Since the algorithm is based on Störmer Verlet, it retains second order accuracy and symplecticity. \cite{grubmuller_generalized_1991, tuckerman_molecular_1991_md_with_long_range_forces, tuckerman_reversible_1992, tuckerman_molecular_1991_md_with_disparate}
The algorithm can not only be used for two different time-scales, but can also be extended to an arbitrary number of separations as can bee seen in \cite{leach_molecular_2001}. It can be easily implemented with a nested loop, where the innermost loop integrates the force that requires the smallest time-step and the outermost loop integrates the force that requires the longest time-step. 
An exemplary pseudo algorithm based on the literature \cite{leach_molecular_2001, griebel_numerical_2007} for a two-force decomposition can be seen in Algorithm \ref{algo:rrespa_theo}. Here, a full time-step $\delta t$ is decomposed into $P$ intermediate steps in the inner loop.
\bigbreak
\begin{algorithm}[H]
    \SetKwFunction{KwNot}{not}
    \SetKwFunction{FcheckCutoff}{checkCutoff}
    \SetKwFunction{FCalculateForces}{CalculateForces}
    \SetKwProg{Fn}{Function}{:}{}
    
    \For{i $\leftarrow$ 0 \KwTo $t_{end}/\delta t - 1$}{
        $\tilde v^i \gets v^i + F_1(x^i) \frac{\delta t}{2m}$\;
        \For{j $\leftarrow$ 1 \KwTo P}{
            $x^{i+j/P} \gets x^{i+(j-1)/P} + \frac{\delta t}{P} \tilde v^{i+(j-1)/P} + (\frac{\delta t}{P})^2 \frac{1}{2m} F_0(x^{i+(j-1)/P})$ \;
            $\tilde v^{i+j/P} \gets \tilde v^{i+(j-1)/P} + \frac{1}{2m} \frac{\delta t}{P} (F_0(x^{i+j-1})F_0(x^{i+j}))$\;
        }
        $v^{i+1} \gets \tilde v^{i+1} + F_1(x^{i+1}) \frac{\Delta t}{2m}$\;
    }

    \caption[r-RESPA]{Pseudo algorithm for r-RESPA based on the literature \cite{leach_molecular_2001, griebel_numerical_2007}. Each time step $\delta t$ is decomposed into $P$ sub-steps in which $F_0$ is integrated in the inner loop. For each full step $\delta t$, $F_1$ is integrated. $v^i$ is the velocity of a particle at time-step $i$, $x^i$ is analogously the position at time-step $i$, $m$ describes the mass of the particle and $t_{end}$ the end time of the simulation.}
    \label{algo:rrespa_theo}
\end{algorithm}
\bigbreak
Note: Our actual implementation does not decompose $\delta t$ into $P$ smaller sub-steps, but instead integrates $F_1$ all $\delta t s$ steps where $s$ is called the step-size-factor. The specific implementation that we use for our work can be found in Algorithm \ref{algo:rrespa} of Section \ref{subsec:implementation}.

\section{Related Work}
\label{sec:relwork}
Multiple Time-Stepping (MTS) methods are commonly employed in simulations where processes occur at different frequencies. Deuflhard et al. \cite{deuflhard_computational_1999} classify time integration into fast and slow degrees of freedom, such as bonded (high-frequency) and non-bonded (low-frequency) interactions. Another key criterion is the particle mass: lighter particles require smaller time steps since they move faster, while heavier particles can be integrated with larger time steps \cite{tuckerman_molecular_1991_md_with_disparate}.

Deuflhard et al. \cite{deuflhard_computational_1999} also distinguish between short-range and long-range forces. Short-range forces, which have a stronger influence on particle motion and higher frequencies, should be integrated with smaller time steps. In contrast, long-range forces, while computationally expensive, have a less significant impact on particle movement and can therefore be integrated with larger time steps.

Grubmüller et al. \cite{grubmller_multiple_1998} compare different MTS methods to reduce computational time for long-range forces. They evaluate a simple cutoff algorithm against distance-class algorithms, which classify particles based on distance and apply the MTS scheme accordingly. Their results indicate that the simple cutoff algorithm is more stable than the MTS methods in terms of energy drift.

The above works (Deuflhard et al. \cite{deuflhard_computational_1999} and Grubmüller et al. \cite{grubmller_multiple_1998}) primarily focus on the application of MTS methods to improve time integration stability and efficiency. However, they do not address the challenges of implementing these methods in the context of High-Performance Computing. Their contributions are largely methodological, with limited consideration for HPC performance factors such as scalability, communication costs, or modern algorithmic optimizations.

Nakano et al. \cite{nakano_parallel_1993} applied MTS methods to reduce communication overhead in three-body molecular dynamics simulations by introducing an intermediate cutoff $r_a$. Three-body and two-body interactions within $0$ to $r_a$ were computed at every time step, while two-body interactions between $r_a$ and $r_c$ were computed less frequently. In their parallel algorithm for distributed memory systems, a speedup of $6.4$ was achieved in a $5184$-particle system with $r_c = 5.5 \mathring{A}$ and $r_a = 2.6 \mathring{A}$, using a time-step factor of $15$. A further parallelized algorithm, which separated the three-body potential calculation, yielded a speedup of $10.7$. The primary focus of this work was to reduce communication costs in the distributed memory algorithm, with limited emphasis on the accuracy of the resulting simulations.

The work by Nakano et al. \cite{nakano_parallel_1993} is the only study we identified that uses MTS methods to accelerate two-body and three-body simulations. However, it fundamentally differs from our approach, as their algorithm computes all three-body interactions at every time step. Moreover, their method is tailored to a distributed memory HPC system from 1993 (Intel iPSC/860) and is of limited relevance for modern HPC architectures. Their primary concern was to mitigate the high communication costs associated with distributed memory algorithms, which they addressed using MTS methods. In contrast, contemporary hybrid cutoff algorithms, where most communication within the cutoff distance occurs in shared memory, are expected to reduce communication costs significantly. For such systems, the computational cost of interactions is often the dominant factor, rather than communication overhead.

Our work takes a novel approach by examining MTS methods specifically in the context of modern HPC systems. We investigate their impact on both performance and accuracy when applied to state-of-the-art algorithms. In particular, we explore the complete decoupling of three-body and two-body interactions using MTS and implement this separation in a distributed memory algorithm optimized for current hardware and MPI communication standards. A key contribution of our study is the development of a novel state-of-the-art cutoff algorithm for three-body interactions. This algorithm is optimized for node-level performance on HPC systems and designed to support hybrid configurations that combine shared and distributed memory. Our evaluation focuses not only on performance improvements but also on the accuracy of the simulations, establishing the potential of MTS methods in modern HPC frameworks.

\section{Evaluation}
\label{sec:evaluation}
In this section we first give a brief insight into the implementation of the algorithms used for three- and two-body interactions, as well as the r-RESPA implementation. Furthermore, we present measured metrics and the test setup. Lastly, we discuss the results of our study.
\subsection{Implementation and Setup}
\label{subsec:implementation}

\subsubruninhead{Direct Sum Implementation}
In a Direct Sum calculation, all interactions between all particles are calculated without cutoff distance.
To calculate all forces between all pairs of particles using the symmetry of Newton's third law, we need to calculate $\binom{N+2-1}{2}$ pairs in a system with $N$ particles. Similarly, we need to calculate $\binom{N+3-1}{3}$ triplets.
Although this is computationally very intensive, it provides fairly accurate results. In the context of our project, this is particularly necessary in order to be able to investigate the effects of different step-size-factors for the three-body interactions. Due to the high computational effort, we opted for a distributed memory implementation which is based on the Algorithm of \cite{koanantakool_computation-_2014}. Based on our literature review, we concluded that this algorithm is currently a state-of-the-art algorithm for the calculation of three-body interactions on HPC clusters. In addition, this algorithm was chosen because, unlike many other algorithms, the calculation of two-body forces can be seamlessly integrated into the calculation of three-body interactions.

\begin{figure}[htbp]
    \begin{subfigure}{.328\textwidth}
        \centering
        \includegraphics[width=\linewidth]{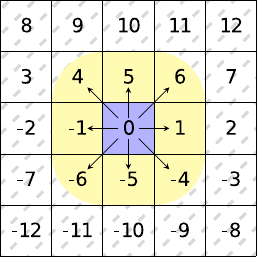}
        \caption{}
        \label{fig:c01a}
    \end{subfigure}
    \begin{subfigure}{.328\textwidth}
        \centering
        \includegraphics[width=\linewidth]{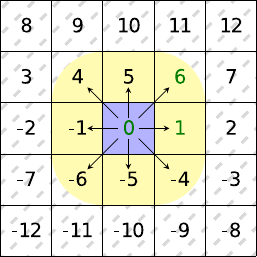}
        \caption{}
        \label{fig:c01b}
    \end{subfigure}
    \begin{subfigure}{.328\textwidth}
        \centering
        \includegraphics[width=\linewidth]{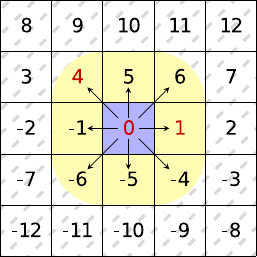}
        \caption{}
        \label{fig:c01c}
    \end{subfigure}
\caption{Visualization of our C01 algorithm for three-body interactions. The simulation domain is divided into a grid, with particles from one or more cells assigned to each processor. The yellow area shows the interaction length based on the cutoff distance within which particles from the central blue cell can interact. Figure \ref{fig:c01b} shows a valid offset pattern consisting of three cells, since starting from each of the cells 0, 1 or 6, the other of the three cells is within the interaction length. Figure \ref{fig:c01c} shows an invalid pattern, as cell 1 is not in the interaction range of cell 4 and vice versa. Triplet interactions are only calculated if the distance between all three particles is less than or equal to the cutoff distance $r_c$.}
\label{fig:c01}
\end{figure}
\subsubruninhead{Cutoff Implementation}
For the \textit{Cutoff} implementation, we used a linked cells algorithm for two-body and three-body interactions. In this algorithm, the particles in the simulation domain are assigned to cells based on their position as illustrated in Figure \ref{fig:c01}. To calculate the interactions between particles, we iterate over the cells and their particles. This can be implemented in different ways and is referred to as traversal. 
\bigbreak
One of the simplest traversals is the C01 traversal by iterating over all cells of the domain, which we call base cells, and calculating the interactions between the particles contained in them and the particles from the neighboring cells that are overlapped by the cutoff distance. The yellow colored area in Figure \ref{fig:c01} represents the area in which particles from the middle blue base cell can interact based on the cutoff distance. We call this interaction length. 
\bigbreak
For two-body interactions, the C01 traversal is easy to implement because either both particles come from the base cell, or one particle comes from the base cell and one from one of the surrounding cells.
\bigbreak
For three-body interactions, either all three particles can come from the base cell, two particles can come from the same cell, or all three particles can come from different cells. In our three-body C01 implementation, we calculate an offset list based on the cutoff distance, which for a base cell contains two neighboring cells per offset entry, with whose particles interactions are calculated. The distance between all three particles must be less than or equal to the cutoff distance so that an interaction is calculated. Figure \ref{fig:c01b} shows a valid offset pattern $(c_1, c_2) = (1, 6)$ starting from a base cell 0. Figure \ref{fig:c01c} shows an invalid offset pattern because cell 4 is not in the interaction radius of cell 1 and vice versa. To calculate only unique offset patterns, we make sure that $c_1 \leq c_2$ lexigographically.
\bigbreak
Since in this traversal each processor processes a base cell and accesses the surrounding cells in all directions, Newton's Third Law of motion cannot be used here.
\bigbreak
This algorithm was implemented in the node level auto tuning software AutoPas, which is a software library for efficient short-range particle simulations with a cutoff distance, like MD simulations. It automatically selects the best algorithm through autotuning during the simulation and is designed for node-level HPC systems. \cite{gratl_n_2022}

\subsubruninhead{r-RESPA Implementation}
The r-RESPA method was implemented uniformly for both applications. In contrast to the approach described in Algorithm \ref{algo:rrespa_theo}, which uses a double-nested loop structure, we simplified it to a single loop where the three-body force is only integrated at full time steps. In the literature, the outer loop handles the integration of the low frequency forces, while the inner loop integrates the high frequency forces at smaller sub-steps. In our implementation, the two-body forces are integrated at each step, while the three-body forces are only integrated every step-size-factor steps. The implementations ensure that the total number of iterations is a multiple of the step-size-factor.
Algorithm \ref{algo:rrespa} gives a rough overview of how r-RESPA time integration is implemented in our case.
\bigbreak
\begin{algorithm}[H]
    \SetKwFunction{KwNot}{not}
    \SetKwFunction{FcheckCutoff}{checkCutoff}
    \SetKwFunction{FCalculateForces}{CalculateForces}
    \SetKwProg{Fn}{Function}{:}{}
    \KwIn{\tabto{2cm}Initial particle positions $x^0$ and velocities $v^0$}
    \KwOut{\tabto{2cm}Particle positions and velocities updated in place.}
    \BlankLine
    $calculate \: F_{2b}(x^0)$\;
    $calculate \: F_{3b}(x^0)$\;
    $s \gets stepSizeFactor$\;
    \For{i $\leftarrow$ 0 \KwTo numIterations - 1}{
    $IsRespaIteration \gets (i \: \% \: s) == 0$\;
    $nextIterationIsRespa \gets ((i+1) \: \% \: s) == 0$\;
    \If{$IsRespaIteration$}{
    $v^i \gets v^i + \frac{\delta t \: s}{2m}F_{3b}(x^i)$\;
    }
    $x^{i+1} \gets x^{i} + \delta t \: v^i + \frac{\delta t^2}{2 m}F_{2b}(x^i)$\; \label{lst:line:pos}
    $calculate \: F_{2b}(x^{i+1})$\; \label{lst:line:f2b}
    $v^{i+1} \gets v^{i} + \frac{\delta t}{2 m}  (F_{2b}(x^i) + F_{2b}(x^{i+1})) $\; \label{lst:line:velupdate}
    \If{$nextIterationIsRespa$}{
    $calculate \: F_{3b}(x^{i+1})$\;
    $v^{i+1} \gets v^{i+1} + \frac{\delta t \: s}{2m}F_{3b}(x^{i+1})$\;
    }
    }

    \caption[r-RESPA]{r-RESPA implementation for two-body and three-body interactions. Note: Only an iteration that is divisible by the step-size-factor can be seen as a full iteration.}
    \label{algo:rrespa}
\end{algorithm}

\subsection{Results}
\label{subsec:results}
In this section, we investigate the r-RESPA method for simulations with two- and three-body interactions, where the two-body interactions are integrated at every time step and the three-body interactions only every few time steps according to the chosen step-size-factor.
\bigbreak
It is worth noting that both the \textit{\textit{DirectSum}} and \textit{Cutoff} methods have their strengths, depending on the desired trade-off between accuracy and computational speed. The \textit{\textit{DirectSum}} method provides higher accuracy as it calculates all interactions without introducing approximations, making it more suitable for analyzing properties like the influence of different step-size-factors on total energy. This is because the \textit{Cutoff} method inherently introduces an approximation due to the truncation of interactions.

For properties like the radial distribution function or pressure, the \textit{Cutoff} method is more practical in our current implementation, as it includes necessary features like boundary conditions, which are not yet available in our \textit{\textit{DirectSum}} implementation. 

However, in principle, boundary conditions could be integrated into \textit{\textit{DirectSum}}, and \textit{Cutoff} implementations could be designed to better account for their effects on total energy.

Ultimately, the choice of method should reflect the specific needs of the simulation. For highly accurate studies where computational cost is secondary, \textit{\textit{DirectSum}} may be preferable. For large-scale or time-sensitive simulations, \textit{Cutoff} offers a more feasible alternative, balancing speed and accuracy effectively.

We use two different scenarios for our measurements. On the one hand, a scenario without reference to a physically correct simulation, but designed to investigate the effects of different $\nu_{ATM}$ parameters for three-body interactions in combination with varying step size factors. On the other hand, a scenario with parameters for a physically correct aluminum simulation. 

In many molecular dynamics simulations, the Lennard-Jones parameters $\epsilon_{LJ}$ and $\sigma_{LJ}$ are scaled to 1, which simplifies the equations and makes them dimensionless. All other parameters, such as the time step $\delta t$, the temperature $T$ or the ATM parameter $\nu_{ATM}$, are then scaled accordingly to maintain consistency. In our aluminum simulation, we use this approach, which is also described in section 3.7.3 of \cite{griebel_numerical_2007}. In all our measurements, the MD systems were first brought into equilibrium before measuring the variables.
\begin{itemize}
    \item \textbf{Toy scenario}: In this scenario we set $\epsilon_{LJ}$ and $\sigma_{LJ}$ to 1 and $\nu_{ATM}$ is varied systematically. This way we simulate a wide range of potential material behaviors and provides a broad overview of the effects of the r-RESPA method under conditions representative of different materials, enabling insights into its applicability and performance across diverse systems. In the scenario, 675 particles were simulated in a $10\times 10 \times 10$ domain over 24000 time-steps with $\delta t = 0.001$.
    \item \textbf{Aluminium:} To investigate the effects of the r-RESPA algorithm on a physically related simulation of a solid, we used the parameters of Branco and Cheng \cite{branco_employing_2021}. The aluminium system with 4995 particles was simulated at $300K$. The corresponding normalized and real values for the simulation can be found in the Appendix in Table \ref{tab:aluminiumparameters}
\end{itemize}

In our measurements, we focus primarily on the accuracy of the simulations with different step-size-factors. However, we also examine the performance. Following metrics have been used to measure the accuracy of the simulations with different step-size-factors:

\begin{itemize}
    \item \textbf{Relative Variation in True Energy (RVITE):} This metric is calculated as
          \begin{align}
              \text{RVITE} = \frac{1}{KJ}\sum_{i=1}^{J} \lvert e(i)-\overline{e} \rvert
          \end{align}
          where $\overline{e}$ is the average total energy, $e(i)$ is the total energy of iteration $i$. $J$ is the number of time steps. $K$ is the average kinetic energy for the simulation. According to Izaguirre et. al. \cite{izaguirre_longer_1999}, the RVITE is a good indicator for the accuracy of symplectic integrators. The smaller this value is, the smaller the RVITE and the more accurate the time integration is accordingly.
          It should be noted that we only use this metric for our measurements with the \textit{\textit{DirectSum}} algorithm. Our results have shown that the \textit{Cutoff} implementation, introduces fluctuations in the total energy. This phenomenon is also described in \cite{diem_effect_2020,feller_effect_1996}. These fluctuations likely obscure the impact of different step-size-factors of the r-RESPA algorithm on this metric, making the effects difficult to measure. For instance, the r-RESPA algorithm showed arbitrary difference in RVITE values when using the \textit{Cutoff} implementation. We suspect that the energy fluctuations caused by the cutoff dominate the metric, rendering the influence of the r-RESPA step-size-factor negligible.
    \item \textbf{Energy Deviation:} Here we compare the values for the total energy with different step-size-factors relative to the total energy measured with Störmer Verlet.
    \item \textbf{Radial distribution function (RDF):} This metric gives us the probability of finding a neighboring particle starting from any arbitrary particle at a certain distance. In our case, this metric helps us to investigate how the distribution of particles in the MD system changes using different step-size-factors.
    \item \textbf{Pressure:} Similar to the RDF, the pressure in an MD system gives us information on the deviation of the particle distribution using different step-size-factors.
    \item \textbf{Performance:} To measure the performance we calculate the speedup based on the total simulation time compared to a step-size-factor of 1 which corresponds to the Störmer Verlet algorithm.
\end{itemize}

\bigbreak

\subsubruninhead{HPC Environment}
All our experiments were carried out on the HSUper compute cluster of the Helmut Schmidt University in Hamburg \footnote{\url{https://www.hsu-hh.de/hpc/en/hsuper/}}. This HPC system has 571 compute nodes, each with 2 Intel(R) Xeon (R) Platinum 8360Y processors with 36 cores each. Thus 72 cores per node.

In all experiments with the \textit{\textit{DirectSum}} implementation, 5 compute nodes with 72 cores each, i.e. a total of 360 cores, were used. For the \textit{Cutoff} algorithm we only used a single node with 72 Hardware Threads.

\subsubsection{Toy Scenario}
In this experiment we investigate the effect of different values for the ATM parameter $\nu_{ATM}$ from $0.05$ to $1.0$ in combination with different step-size-factors from 1 to 12 on the Relative Variation in True Energy. The parameters for the ATM potential were chosen to cover a wide range from gaseous to solid materials.

\begin{figure}[htb]
\centering
    \includegraphics[width=\linewidth]{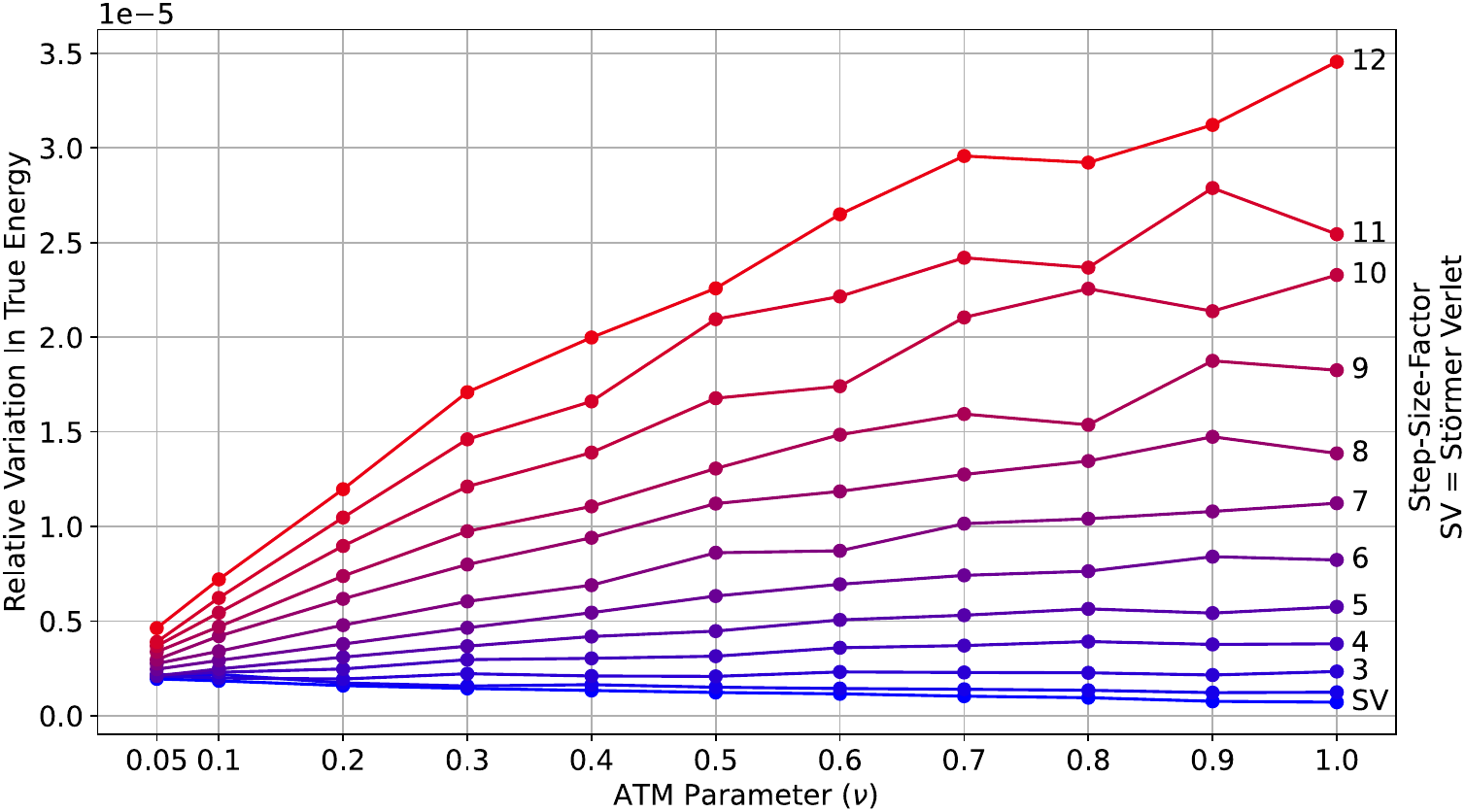}
\caption{Comparison of the Relative Variation In True Energy for different step-size-factors and different values for the $\nu_{ATM}$ parameter. Each line represents one step-size-factor. From bottom to top, this corresponds to a step-size-factor of 1 (Störmer Verlet) up to 12.}
\label{fig:rvite}
\end{figure}

\bigbreak
As we can see in Figure \ref{fig:rvite}, there is a decreasing trend in RVITE for a step-size-factor of 1 and 2 as the $\nu_{ATM}$ parameter increases. This downward trend suggests that the inclusion of three-body interactions in combination with Störmer-Verlet integration contributes to a more stable behavior of the system, particularly as the $\nu_{ATM}$ parameter grows. For this reason, we assume that for these step-size-factors, increasing the $\nu_{ATM}$ parameter has a stronger influence on the overall simulation. For a step-size-factor of 3, we see a similar value for all $\nu_{ATM}$ parameters. For the higher step-size-factors from 4 to 12, we then see an increase in the RVITE value as the $\nu_{ATM}$ parameter increases and thus a loss of accuracy of the r-RESPA algorithm.
Our literature research has shown that lower values for the $\nu_{ATM}$ parameter are usually used for gaseous materials, which in Figure \ref{fig:rvite} would rather be in the range $0.2$ to $0.3$ \cite{akhouri_thermodynamic_2022}. For solid materials, such as our aluminum simulation, higher values in the range $0.7$ are used \cite{branco_employing_2021}. We can therefore conclude that the stability depends on the material to be simulated and that higher step-size-factors are more effective for gaseous materials than for solid materials.

\subsubsection{Aluminium}

\begin{figure}[ht]
\centering
\includegraphics[width=10.5cm]{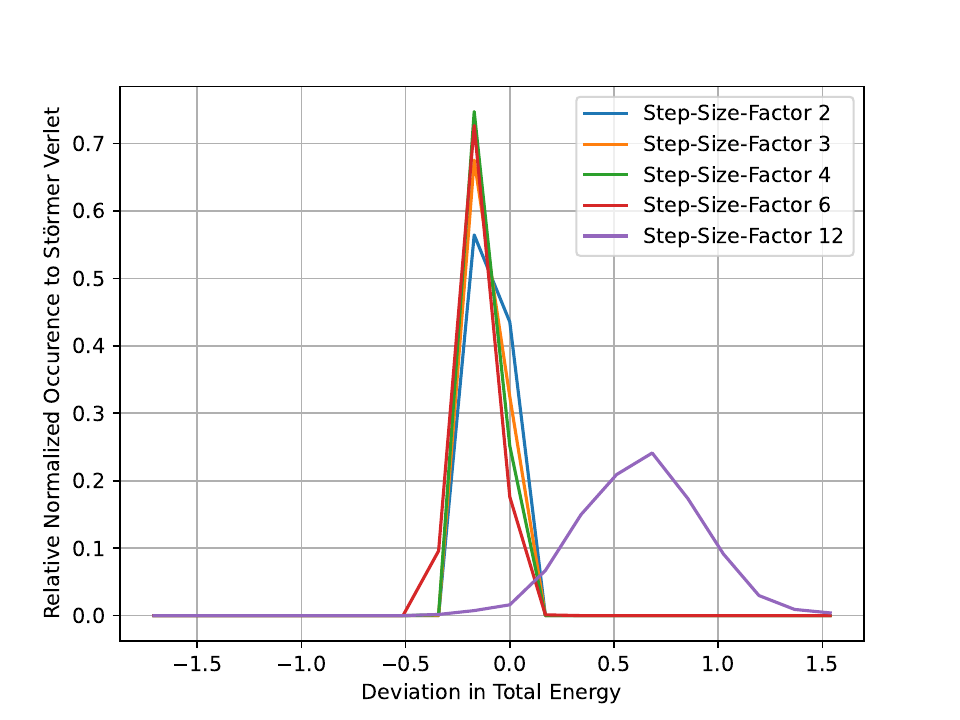}
\caption{Values of total energy relative to Störmer Verlet. Here, the range between the minimum and maximum total energy across all step-size-factors was divided into 50 bins and the values of the respective simulations of each iteration were distributed across these bins. The line for Störmer Verlet was intentionally not shown here. It would be a vertical line at 0.0 with a height of 1.}
\label{fig:energydeviation}
\end{figure}

\bigbreak
In Figure \ref{fig:energydeviation} we have performed a simulation with the aluminum parameters using the \textit{\textit{DirectSum}} algorithm and different step-size-factors for the r-RESPA time integration algorithm. 

The histogram shows the relative deviation in the total energy for different step-size factors compared to an integration of both forces in each time step (Störmer Verlet). In the center at 0.0, the Störmer Verlet algorithm would be seen as a single line with height 1, which is intentionally not shown here. All other lines represent the relative deviations with the corresponding step-size-factors. Only energy values corresponding to every twelfth iteration were used, as this corresponds to a full r-RESPA time-step for all step-size factors. The value of the minimum and maximum energy over the entire simulation was determined for all simulations. This range was divided into 50 bins and the energy values of each simulation were distributed accordingly. This gives us an overview of how the total energy over the entire simulation behaves in relation to Störmer Verlet. 

To put the deviations on the x-axis into context, the mean energy values for Störmer Verlet and a step-size factor of 12 are as follows: For Störmer Verlet, the mean energy was $-16637.88$, while for a step-size factor of 12, it was $-16637.20$. The maximum deviations from the mean were approximately $0.0005\%$ for Störmer Verlet and $0.006\%$ for a step-size factor of 12. This highlights that for increasing step-size factors, the oscillations in the total energy become more pronounced. This also shows that for increasing step-size-factors the oscillations in the total energy increase. 

For all step-sizes from 2 to 6 a slight negative shift of the total energy compared to an integration of both forces in each time step can be seen. For the step-size-factor of 12, a clear shift of the total energy to the right is recognizable. It should be noted that there is a large gap of step-size-factors between step-size-factor 6 and 12, which we could not measure as these values do not have a common divisor for full steps. However, it is clear to see that the stability of the time integration decreases significantly with a step-size-factor of 12 in this case. The directions in which the energy drifts were not investigated further in this case, and it is a matter of ongoing research to explain these drift directions.

\begin{figure}[htbp]
\centering
    \includegraphics[width=10.5cm]{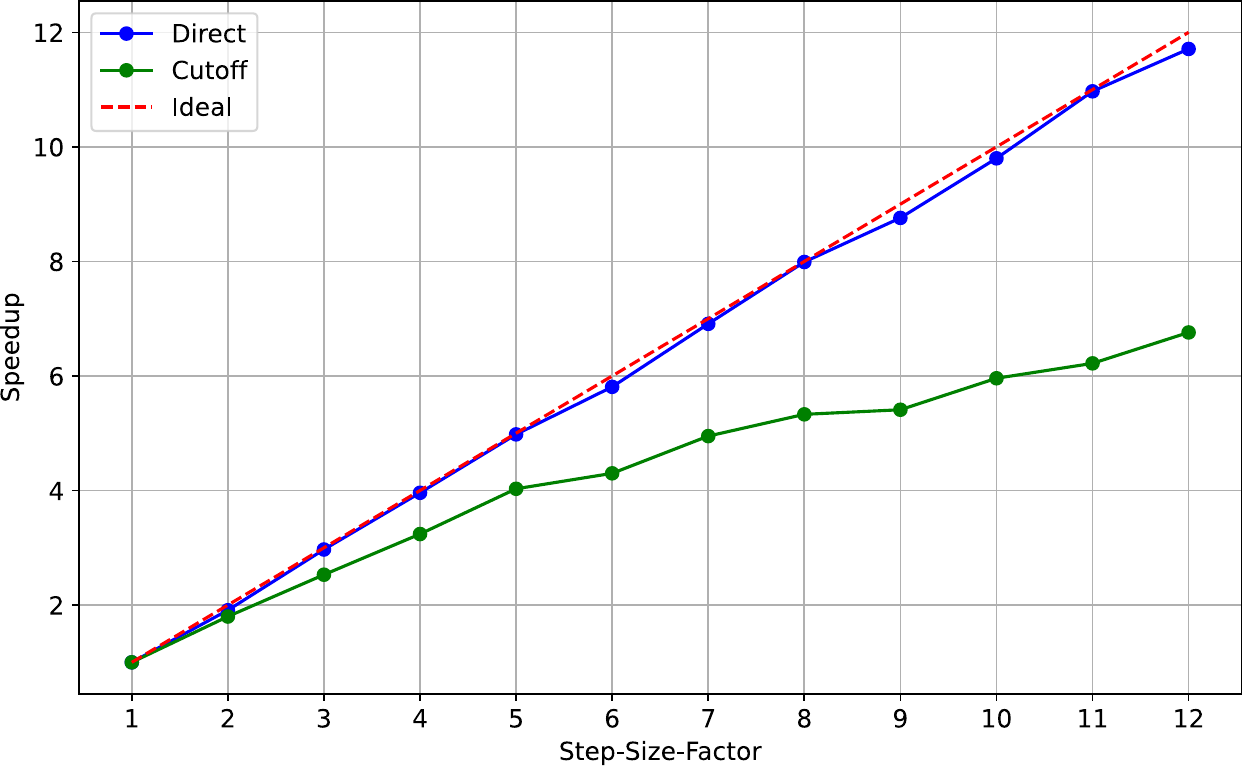}
\caption{Speedup for the aluminum simulation wit the \textit{Cutoff} and the \textit{DirectSum} algorithms. The ideal speedup is shown as a red dashed line.}
\label{fig:speedupdirect}
\end{figure}

\bigbreak
To investigate the effects on performance, we examined the speedup with different step-size-factors in Figure \ref{fig:speedupdirect}. These measurements were carried out using both the \textit{\textit{DirectSum}} algorithm and the \textit{Cutoff} algorithm. The calculation time for the \textit{DirectSum} algorithm is reduced by almost half with a step-size-factor of 2 compared to Störmer Verlet. This trend continues for the other step-size-factors. The result was to be expected, since the three-body interactions take up most of the calculation time compared to the two-body interactions. 

When using the \textit{Cutoff} algorithm, the speedup decreases as the step-size-factor increases. In the direct implementation, all possible triplets and pairs within the simulation domain are calculated. However, in the \textit{Cutoff} implementation, only interactions that meet the cutoff criterion are considered. For pairwise interactions, the distance between two particles must be less than or equal to the cutoff, while for triplets, all three pairwise distances between the three particles involved must be within the cutoff distance. Our measurements for the \textit{Cutoff} algorithm show that a two-body plus three-body simulation with both forces integrated at each step, compared to a pure two-body simulation, requires about 13.4 times more computation time. In contrast, the Direct Algorithm for the same two-body plus three-body simulation takes about 393.6 times more computation time. This illustrates that the ratio of calculated three-body to two-body interactions is significantly lower in the \textit{Cutoff} implementation compared to the \textit{DirectSum} algorithm, leading to a smaller speedup. The reduced dominance of three-body calculations in the \textit{Cutoff} method explains why the speedup is not as significant as in the \textit{DirectSum} algorithm.

\begin{figure}[ht]
\centering
    \includegraphics[width=10.5cm]{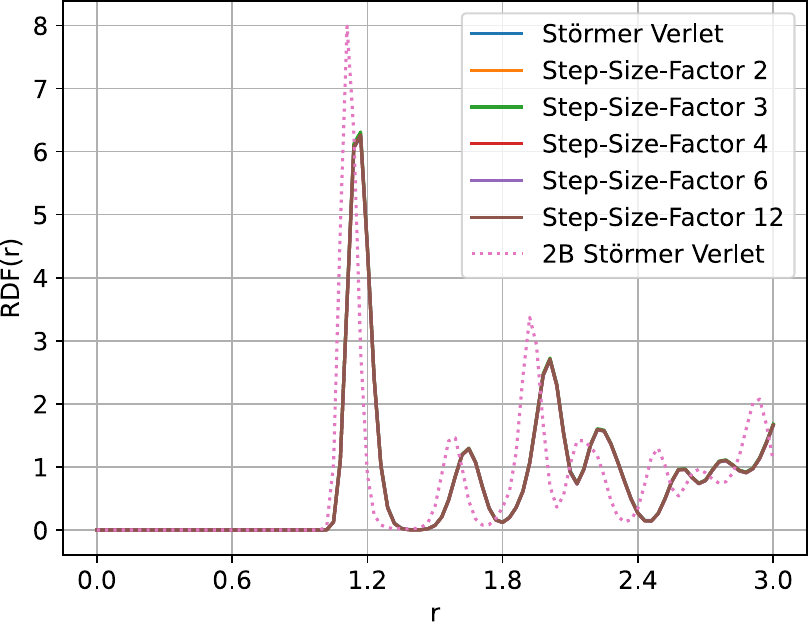}
\caption{Radial Distribution Function for the aluminium simulation including a pure two-body simulation and two-body plus three-body simulations with different step-size-factors for three-body interactions. The \textit{Cutoff} algorithm was used here with periodic boundary conditions.}
\label{fig:rdf}
\end{figure}

\bigbreak
To investigate deviations in the density of the MD system with different step-size-factors for the three-body interactions, we calculated the results of the radial distribution function using the \textit{Cutoff} algorithm and the parameters for aluminum. We performed two-body plus three-body simulations with different step-size-factors. In addition, we performed a pure two-body simulation to demonstrate how the density of an MD system changes when three-body interactions are included. In Figure \ref{fig:rdf} the pure two-body simulation is shown as a dashed line. All other lines are two-body plus three-body simulations with different step-size-factors. We see that there are clear differences in the radial distribution function between a pure two-body and a two-body plus three-body simulation. We conclude that the incorporation of three-body interactions has a noticeable influence on the structure of the system. However, the deviations are comparatively small when using different step size factors for the integration of the three-body interactions.

\begin{figure}[ht]
\centering
    \includegraphics[width=10.5cm]{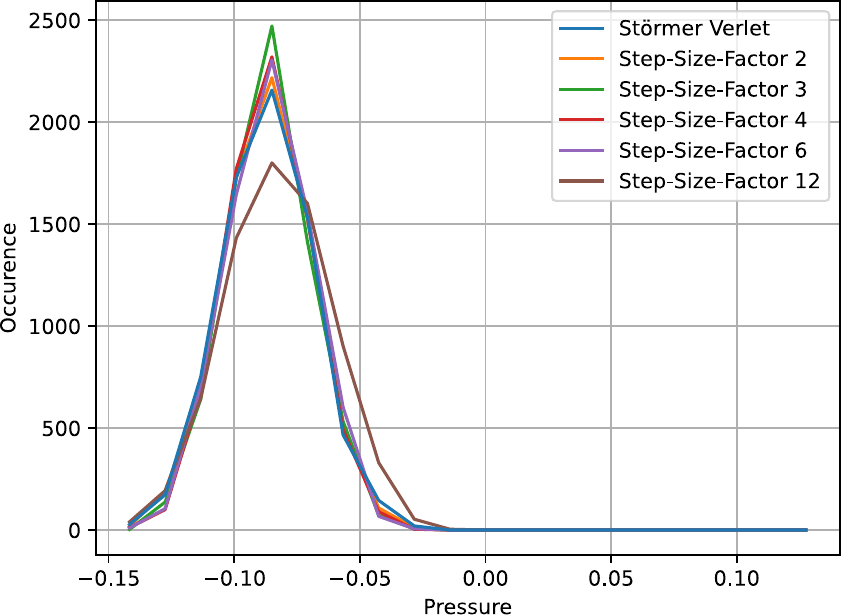}
\caption{Comparison of the absolute values for the pressure of the MD system with different step-size-factors. The \textit{Cutoff} algorithm was used here with periodic boundary conditions. The values of all iterations for the respective step-size-factors were distributed over 20 bins. The pressure values were calculated here over 84000 iterations after reaching the equilibrium.}
\label{fig:pressure}
\end{figure}

\bigbreak
In Figure \ref{fig:pressure} we have examined the pressure for the aluminum simulation using the \textit{Cutoff} algorithm with periodic boundary conditions. For the step size factors 2 to 6, no significant shift in the values compared to Störmer Verlet can be seen. A pressure shift can be seen here for a step size factor of 12. The peak value, however, remains at the same level as for the other step size factors.

\section{Conclusion}
\label{sec:conclusion}

In our work, we investigated the use of the r-RESPA method to optimize the computation time of three-body interactions in simulations involving both two-body and three-body interactions. By integrating three-body interactions with larger time steps, we aimed to achieve significant speedups while maintaining the accuracy of key physical properties.

We evaluated this approach through three scenarios: a toy-scenario that, while not based on a physical simulation, encompasses parameters commonly encountered in real systems, and one real-world scenario with parameters for aluminum. These scenarios were analyzed using our \textit{DirectSum} and \textit{Cutoff} implementations, with step-size factors varied to measure deviations in total energy and structural properties like the radial distribution function and pressure.

Our results demonstrate that the r-RESPA method can be used to improve the performance of such MD simulation significantly. The accuracy of the resulting quantities depends on the materials to be simulated. For simulations that use the Axilrod-Teller-Muto Potential for three-body interactions, it has been found that materials in which the three-body potential has less influence also maintain stability in the total energy better with larger step-size factors. This includes gases, for example. For solid materials, such as our aluminum simulation, instability grows faster with increasing step-size factors. However, it was also shown that the density is maintained in a simulation with a solid for almost all step-size factors.

Looking ahead, it would be also interesting to explore the concept of different distance classes as introduced in \cite{grubmller_multiple_1998,nakano_parallel_1993}. In this approach, the cutoff distance is divided into multiple intermediate cutoffs, with interactions integrated at varying time steps based on their distances. By applying this idea to three-body interactions, one could investigate whether calculation times can be further optimized while preserving the accuracy of measured quantities.

In order to further improve the performance of three-body interactions in cutoff algorithms, more efficient traversals could be developed in addition to the C01 traversal, which, for example, also allow the use of Newton's Third law of motion.

More efficient traversals in combination with the distance classes and an optimization of tuning parameters in Autopas could further increase the performance of such simulations.

\section{Acknowledgements}
\label{sec:acknowledgements}
Computational resources (HPC cluster HSUper) have been provided by the project hpc.bw, funded by dtec.bw – Digitalization and Technology Research Center of the Bundeswehr. dtec.bw is funded by the European Union – NextGenerationEU.
This work has been supported by the project MaST, which is funded by dtec.bw – Digitalisation and Technology Research Center of the Bundeswehr. dtec.bw is funded by the European Union – NextGenerationEU. The authors also gratefully acknowledge funding for the present work by the Federal Ministry of Education and Research (BMBF, Germany), project 16ME0653 (3xa). BMBF is funded by the European Union - NextGenerationEU.

\section*{Appendix}
\addcontentsline{toc}{section}{Appendix}
\label{sec:appendix}

\begin{table}[h]
\caption{Parameters used for the aluminium scenario based on \cite{branco_employing_2021}.}
\label{tab:aluminiumparameters}
\begin{tabular}{p{110.965743333pt}p{110.965743333pt}p{110.965743333pt}}
\hline\noalign{\smallskip}
Property / Parameter & Dimensionless Value & Actual Value \\
\hline
\vspace{2pt}
Domain Size          & 20x20x20                       & 20x20x20                            \\
Number of Particles  & 4995                           & 4995                                \\
Number of Iterations & 24000                          & 24000                               \\
$\delta t$           & 0.00304                        & 1 fs                                \\
$m$                  & 1                              & 26.982u                             \\
$\epsilon_{LJ}$      & 1                              & $2.688\cdot10^{-20}J$               \\
$\sigma_{LJ}$        & 1                              & 2.5487 $\mathring{A}$               \\
$\nu_{ATM}$          & 0.3095                         & $3.776\cdot10^{-17}J\mathring{A}^9$ \\
$T$                  & 1.1                            & $300 K$ \\
\noalign{\smallskip}\hline\noalign{\smallskip}
\end{tabular}
\end{table}

\end{document}